\documentclass[preprint]{aastex}
\usepackage{float}
\usepackage{color}

\title{A Search for $l$-C$_3$H$^+$ and $l$-C$_3$H in Sgr B2(N), Sgr B2(OH), and the Dark Cloud TMC-1}
\author{Brett A. McGuire \& P. Brandon Carroll}
\affil{Division of Chemistry and Chemical Engineering, California Institute of Technology, Pasadena, CA 91125}
\author{Ryan A. Loomis}
\affil{Department of Chemistry, University of Virginia, Charlottesville, VA, 22904}
\author{Geoffrey A. Blake}
\affil{Division of Chemistry and Chemical Engineering and Division of Geological and Planetary Sciences, California Institute of Technology, Pasadena, CA 91125}
\author{Jan M. Hollis}
\affil{NASA Goddard Space Flight Center, Greenbelt, MD, 20771}
\author{Frank J. Lovas}
\affil{National Institute of Standards and Technology, Gaithersburg, MD 20899}
\author{Philip R. Jewell \& Anthony J. Remijan}
\affil{National Radio Astronomy Observatory, Charlottesville, VA 22903}

\begin{document}

\begin{abstract}

Pety et al. (2012) recently reported the detection of several transitions of an unknown carrier in the Horsehead PDR and attribute them to $l$-C$_3$H$^+$.  Here, we have tested the predictive power of their fit by searching for, and identifying, the previously unobserved $J=1-0$ and $J=2-1$ transitions of the unknown carrier (B11244) towards Sgr B2(N) in data from the publicly available PRIMOS project.  Also presented here are observations of the $J=6-5$ and $J=7-6$ transitions towards Sgr B2(N) and Sgr B2(OH) using the Barry E. Turner Legacy Survey and results from the Kaifu et al. (2004) survey of TMC-1.  We calculate an excitation temperature and column density of B11244 of $\sim$10 K and $\sim$$10^{13}$ cm$^{-2}$ in Sgr B2(N) and $\sim$79 K with an upper limit of $\leq1.5 \times 10^{13}$ cm$^{-2}$ in Sgr B2(OH) and find trace evidence for the cation's presence in TMC-1.  Finally, we present spectra of the neutral species in both Sgr B2(N) and TMC-1, and comment on the robustness of the assignment of the detected signals to $l$-C$_3$H$^+$.
\end{abstract}

\section{Introduction}
\label{intro}

The identification and characterization of molecular species in the interstellar medium (ISM) has traditionally followed a linear progression.  Species of interest, perhaps highlighted by chemical models, are obtained or produced in the laboratory, and their characteristic spectra (rotational, vibrational, etc.) are measured.  These spectra are fit to constants unique to each species which can then be used, with knowledge of the Hamiltonian, to reproduce the spectra and predict the appearance of additional features under interstellar conditions.   Observations of either the measured lab features or calculated transitions in the ISM can then be used to unambiguously identify and characterize new molecules in astronomical environments.  

Pety et al. (2012) recently reported the first detection of $l$-C$_3$H$^+$ in the ISM using an IRAM 30 m line survey of the Horsehead PDR.  An extensive search of the literature has uncovered no prior laboratory work to characterize the rotational spectra or constants of $l$-C$_3$H$^+$.  This identification is therefore significant - the number of molecular species detected in the ISM via rotational transitions without the \textit{a priori} knowledge of laboratory spectra or constants is quite small.   Notable among these is the detection of the HCO$^+$ ion, popularly attributed to unidentified features in observations by Buhl and Snyder in 1970 \citep{Buhl1970} dubbed ``xogen," but not definitively detected until laboratory measurements were available nearly six years later \citep{Woods1975} following the suggested theoretical assignment of Klemperer (1970).  The N$_2$H$^+$ ion in 1975 \citep{Green1974,Thaddeus1975} following observation in 1974 \citep{Turner1974} which was confirmed by laboratory studies in 1976 \citep{Saykally1976}.  Shortly thereafter, Gu\'{e}lin, Green, \& Thaddeus identified the C$_3$N \citep{Guelin1977} and C$_4$H \citep{Guelin1978} radicals based on their observations which were confirmed by theoretical studies \citep{Wilson1977} and laboratory measurements several years later \citep{Gottlieb1983}.  More recently, strong evidence for the detection of the C$_5$N$^-$ anion has been found towards IRC+10216 in work by Cernicharo et al. (2008), supported by the \textit{ab initio} calculations of Botschwina et al. (2008).

In the case of $l$-C$_3$H$^+$, the simplicity of the rotational spectrum of a ground state, closed-shell, linear molecule enabled Pety and coworkers to make a compelling case for its detection based on eight rotational transitions observed in their survey.  As described in Pety et al. (2012), $l$-C$_3$H$^+$ provides a valuable probe into small hydrocarbon chemistry in the ISM.  Its characterization in observations using state-of-the-art facilities such as Herschel and ALMA is therefore desirable.

However, the veracity of the assignment of these signals to $l$-C$_3$H$^+$ was questioned by Huang and coworkers \citep{Huang2013}.  They performed high-level quantum chemical calculations to determine rotational and distortion constants for the molecule.  They find large discrepancies between their calculated values for the $D$ and $H$ distortion constants and those determined by Pety et al. (2012); several orders of magnitude in the case of $H$.  As a result, they question the assignment of the signals to $l$-C$_3$H$^+$.  In this work, we will assume the carrier of the signals detected here and by Pety et al. (2012) is a closed-shell linear molecule and refer to it as B11244. In light of the work by Huang et al. (2013), here we present observations of the neutral $l$-C$_3$H molecule as well as B11244 in three new sources, providing additional context for the debate.\footnote{Note that during the final revision of this manuscript, further theoretical work toward the identification of B11244 was performed by Fortenberry et al. (2013) and released on the arXiv.  Their work suggests the C$_3$H$^-$ anion as a candidate for the identity of B11244.  We encourage the interested reader to review this work, but do not discuss it further here.}

\section{Observations}
\label{obs}

The cm-wave data presented here towards Sgr B2(N) were taken as part of the \textbf{PR}ebiotic \textbf{I}nterstellar \textbf{MO}lecular \textbf{S}urvey (PRIMOS) project using the National Radio Astronomy Observatory's (NRAO) 100-m Robert C. Byrd Green Bank Telescope.  The PRIMOS key project began in January of 2008 and observations continue to expand its frequency coverage.  This project provides high-resolution, high-sensitivity spectra of the Sgr B2(N-LMH) complex centered at (J2000) $\alpha$ = 17h47m110s, $\delta$ = -28$^{\circ}$22\arcmin17$\arcsec$ with nearly continuous frequency coverage from 1 - 50 GHz.  The 2 mm observations (hereafter the Turner Survey) were conduced by Barry E. Turner using the NRAO 12-m Telescope on Kitt Peak between 1993 - 1995 towards a number of sources, including Sgr B2(N) and the associated Sgr B2(OH).  A complete description of the PRIMOS observations is given in Neill et al. (2012).\footnote{Specifics on the observing strategy including the overall frequency coverage and other details for PRIMOS are available at http://www.cv.nrao.edu/$\sim$aremijan/PRIMOS/.}  Details of the Turner Survey can be found in Pulliam et al. (2012) or in Remijan et al. (2008).\footnote{Access to the entire PRIMOS dataset and complete Turner Survey are available at http://www.cv.nrao.edu/$\sim$aremijan/SLiSE.}  We also present observations of TMC-1, the details of which are given in Kaifu et al. (2004). The data are reproduced here with permission.

\begin{figure}
\plotone{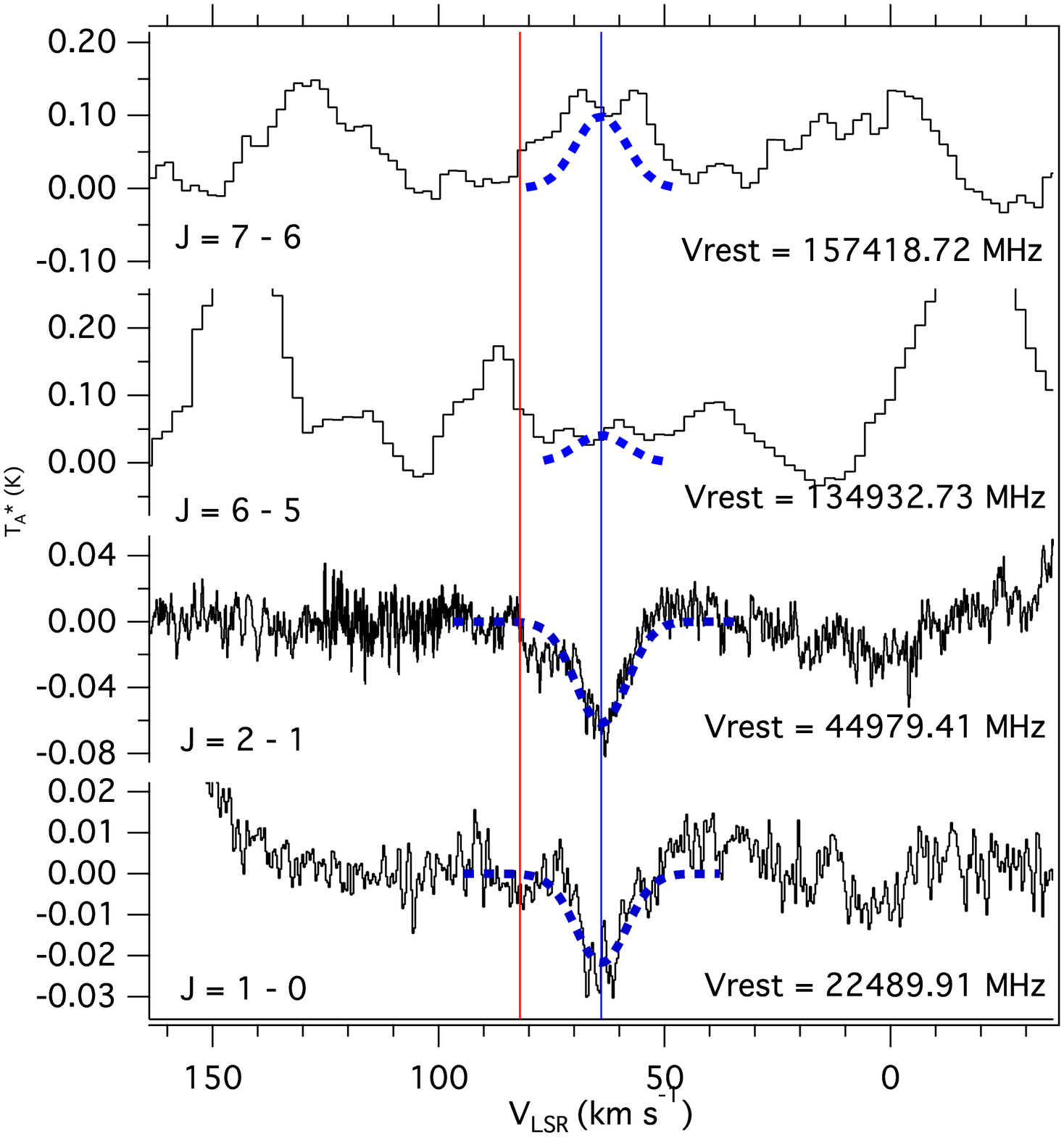}
\caption{Observed transitions of B11244 towards Sgr B2(N).  Plots are on a common velocity scale, with rest frequencies assuming a V$_{LSR}$ = +64 km s$^{-1}$ and line centers taken as those fitted by Pety et al. (2012). Blue and red lines indicate the +64 and +82 km s$^{-1}$ common velocity components in observations of Sgr B2, respectively.  Predictions of line profiles and intensities in the Sgr B2(N) observations based on the best fit temperature and column density determined from the $J=1-0$ and $J=2-1$ transitions are shown as a dashed profile in blue.}
\label{b2n}
\end{figure}

\begin{figure}
\plotone{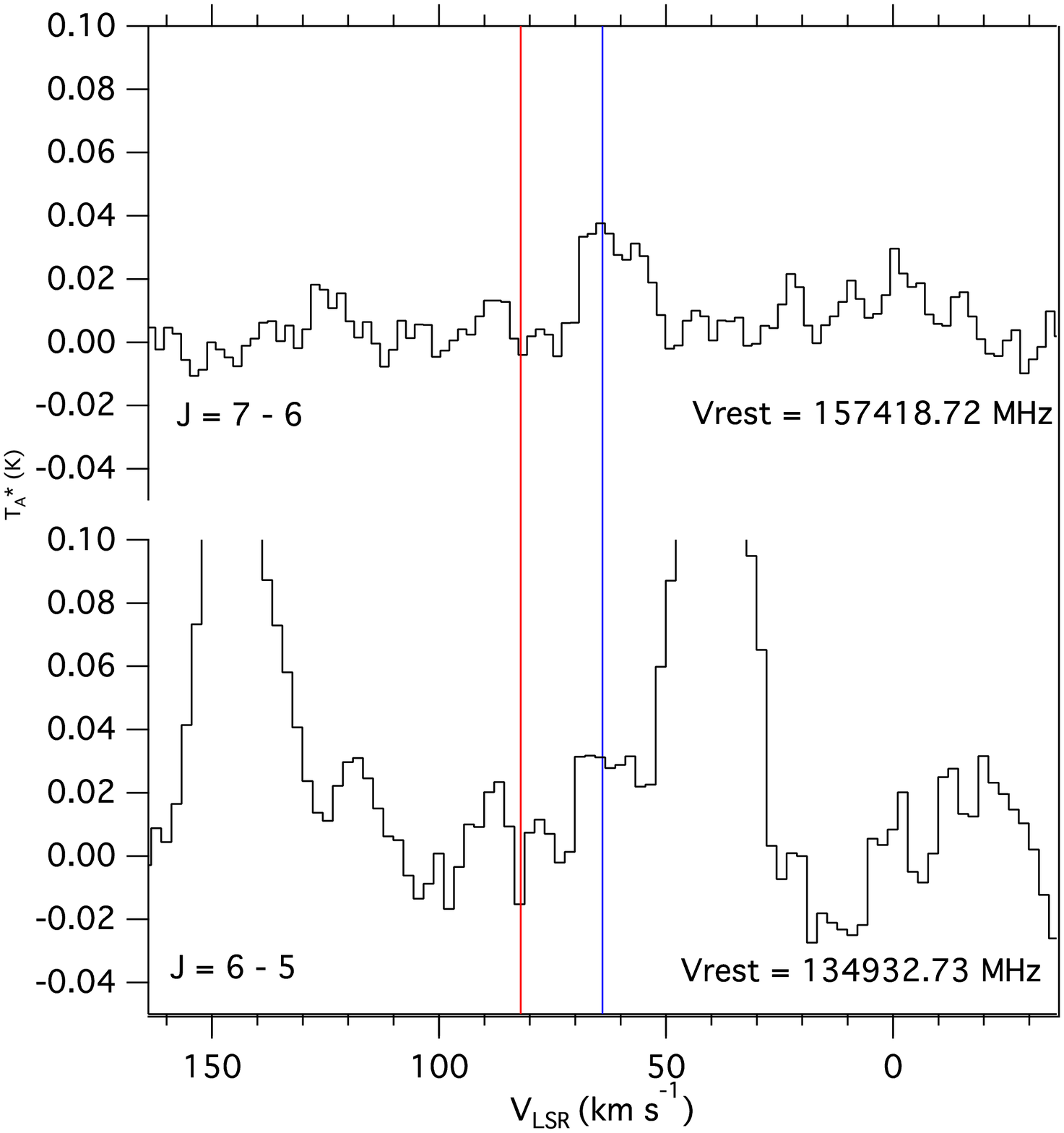}
\caption{Observed transitions of B11244 towards Sgr B2(OH).  Plots are on a common velocity scale, with rest frequencies assuming a V$_{LSR}$ = +64 km s$^{-1}$ and line centers taken as those fitted by Pety et al. (2012). Blue and red lines indicate the +64 and +82 km s$^{-1}$ common velocity components in observations of Sgr B2, respectively.}
\label{b2oh}
\end{figure}

\subsection{Sgr B2(N)}

The observed $J=1-0$ and $J=2-1$ transitions of B11244 towards Sgr B2(N) are shown in Figure \ref{b2n}, with each spectral region shifted to the rest frequency as predicted by Pety et al. (2012) and assuming a $V_{LSR} = +64$ km s$^{-1}$.  There are clear absorption signals for both transitions at the +64 km s$^{-1}$ component.  A less intense, but visible, absorption feature is observed in the +82 km s$^{-1}$ component for the  $J=1-0$ line.  This is consistent with previous observations of molecular signals in this source (e.g. HNCNH, McGuire et al. (2012)), where much weaker signals are observed in the +82 km s$^{-1}$ component.  At higher frequencies from the Turner Survey, weak emission is seen at the frequencies of the $J=6-5$ and $J=7-6$ transitions.  The observed intensities of these features are consistent with the column densities and temperatures derived in \S\ref{data} and provide a constraint on the continuum temperature at these frequencies.  The observed intensities and linewidths are given in Table \ref{observedtransitions}.  

To further confirm the detection, we have performed an analysis of the probability of coincidental overlap of the $J=1-0$ and $J=2-1$ transitions following the convention of Neill et al. (2012).  Using the parameters from Neill et al. (2012) for the line density of absorption features and weak absorption features, and assuming a conservative FWHM of 25 km s$^{-1}$, we find the probability of a single coincidental overlap to be $\sim$0.05.  For two coincidental transitions, this probability falls to $\sim$0.003.

\begin{figure}
\plotone{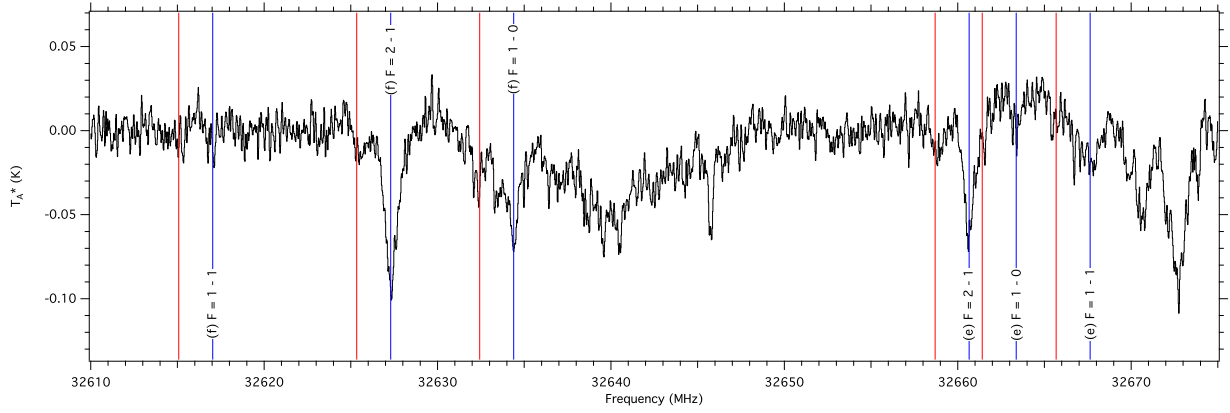}
\caption{The $J = 3/2 - 1/2$, $\Omega = 1/2$ transitions of $l$-C$_3$H toward Sgr B2(N) from PRIMOS.  Rest frequency is adjusted for a V$_{LSR}$ = +64 km s$^{-1}$. Blue and red lines indicate the +64 and +82 km s$^{-1}$ common velocity components in observations of Sgr B2, respectively.}
\label{linear}
\end{figure}

For comparison, we have also searched for the $J = 3/2 - 1/2$ rotational branch of the neutral $l$-C$_3$H molecule occurring around 32.6 GHz.  The PRIMOS spectra are shown in Figure \ref{linear}.  Absorption is clearly observed in all lines at +64 km s$^{-1}$, while weaker absorption is seen in the +82 km s$^{-1}$ component.  The weaker absorption in the  +82 km s$^{-1}$ component compared to the +64 km s$^{-1}$ is consistent with observations of the B11244 species.  The observed intensities and linewidths are given in Table \ref{neutraltransitions}

\subsection{Sgr B2(OH)}

Our coverage of Sgr B2(OH) includes only the $J = 6-5$ and $J = 7-6$ transitions of B11244. In the Sgr B2(OH) complex, these signals are much clearer in the +64 km s$^{-1}$ component than in Sgr B2(N), but no signal is seen in the +82 km s$^{-1}$ component (see Figure \ref{b2oh}).   The lines are observed in emission, and are likely blended with neighboring transitions.  No signal from neutral $l$-C$_3$H is observed in the available data toward Sgr B2(OH).

\subsection{TMC-1}

\begin{figure}
\plotone{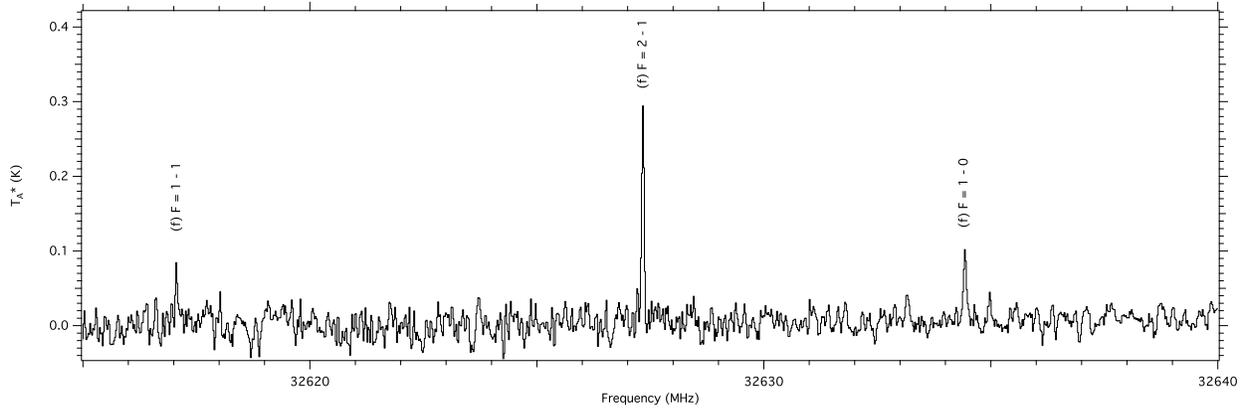}
\caption{The $J = 3/2 - 1/2$, $f$-parity transitions of $l$-C$_3$H toward TMC-1 from Kaifu et al. (2004).  Rest frequency is adjusted for a V$_{LSR}$ = +5.85 km s$^{-1}$}
\label{tmc1a}
\end{figure}

\begin{figure}
\plotone{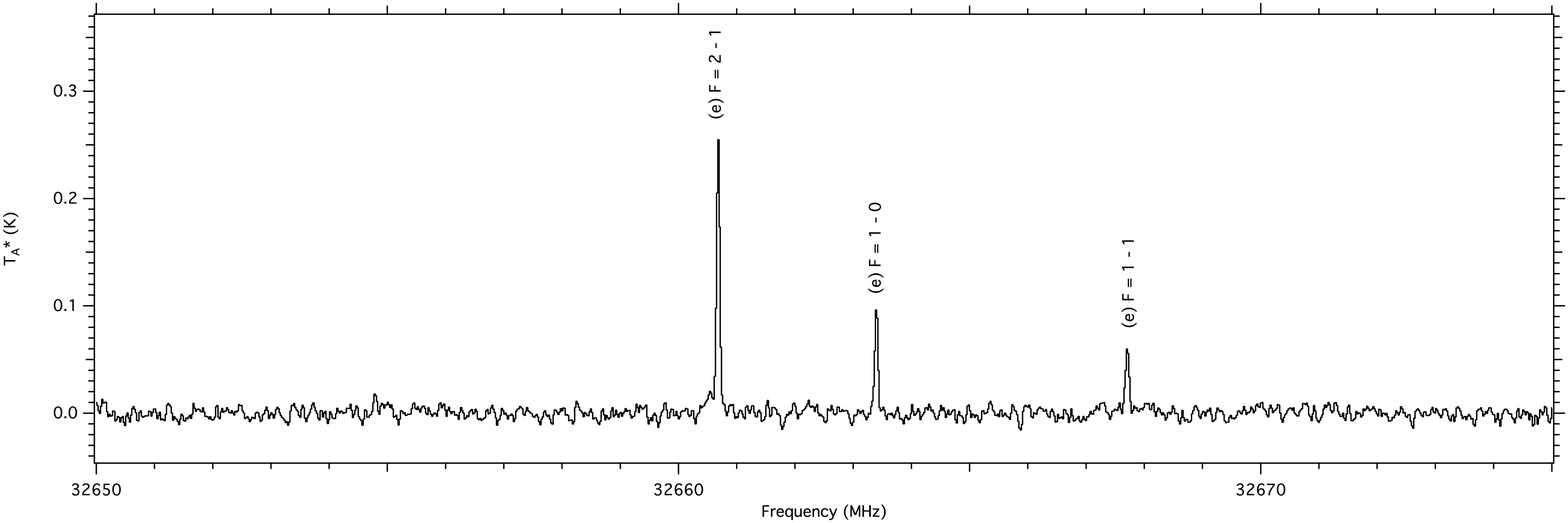}
\caption{The $J = 3/2 - 1/2$, $e$-parity transitions of $l$-C$_3$H toward TMC-1 from Kaifu et al. (2004).  Rest frequency is adjusted for a V$_{LSR}$ = +5.85 km s$^{-1}$}
\label{tmc1b}
\end{figure}

Kaifu et al. (2004) observed the $J = 3/2 - 1/2$ hyperfine transitions of $l$-C$_3$H in their survey of TMC-1, building on a previous detection of the molecule in this source and in IRC+10216 \citep{Thaddeus1985}.  In their survey, Kaifu et al. observed two additional transitions of $l$-C$_3$H, the two weakest $F = 1-1$ hyperfine lines, and used these measurements to further refine the constants originally derived by Thaddeus et al. (1985) and in the laboratory by Gottlieb et al. (1986).  The detected transitions are shown in Figures \ref{tmc1a} and \ref{tmc1b}, and the parameters reported by Kaifu et al. (2004) are given in Table \ref{neutraltransitions}.

\begin{figure}
\plotone{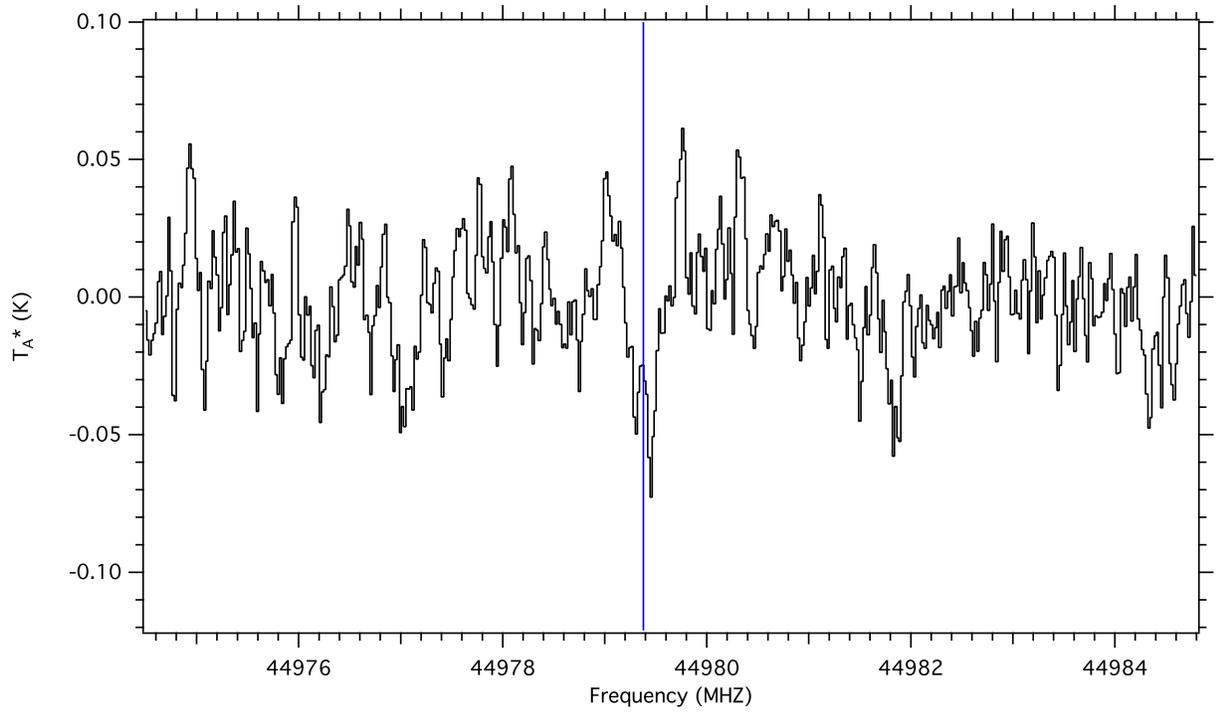}
\caption{The $J = 2-1$ transition of B11244 toward TMC-1 from the Kaifu et al. (2004) data.  Rest frequency is adjusted for a V$_{LSR}$ = +5.85 km s$^{-1}$ and indicated by a blue line.}
\label{tmc1ion}
\end{figure}

No evidence for the $J = 1-0$ transition of B11244 in emission or absorption is present in the TMC-1 data.  Very weak absorption is seen at the frequency of the $J = 2-1$ transition as shown in Figure \ref{tmc1ion}.  While tantalizing, it is certainly not definitive evidence of B11244's presence. 

\begin{deluxetable}{l c c c c c c c c c c c}
\tablewidth{0pt}
\tablecaption{Transitions of B11244 observed toward Sgr B2(N), Sgr B2(OH), and TMC-1.}
\tablecolumns{12}
\tabletypesize{\footnotesize}
\tablehead{
\colhead{} 			& \colhead{} 			& \colhead{} 		& \colhead{} 				& 							\multicolumn{4}{c}{Sgr B2(N)} 									& \multicolumn{2}{c}{Sgr B2(OH)} 					& \multicolumn{2}{c}{TMC-1} 					\\ \cline{5-12}
\colhead{} 			& \colhead{} 			& \colhead{} 		& \colhead{} 				& \multicolumn{2}{c}{+64 km s$^{-1}$} 	& \multicolumn{2}{c}{+82 km s$^{-1}$} 	&  \colhead{} 			&  \colhead{} 				&  \colhead{} 			&  \colhead{} 								\\ \cline {5-8}
\colhead{} 			& \colhead{Frequency} 	& \colhead{E$_u$} 	& \colhead{} 				& \colhead{T$_A^*$}	&  \colhead{$\Delta V$} 	&  \colhead{T$_A^*$} 		&  \colhead{$\Delta V$} 	&  \colhead{T$_A^*$} 		&  \colhead{$\Delta V$} 	&  \colhead{T$_A^*$} 	&  \colhead{$\Delta V$} 	\\
\colhead{Transition}	 	& \colhead{MHz} 		& \colhead{K} 		& \colhead{S$_{ij}\mu^2$} 	& \colhead{mK} 		&  \colhead{km s$^{-1}$} 	&  \colhead{mK}			&  \colhead{km s$^{-1}$} 	&  \colhead{mK} 			&  \colhead{km s$^{-1}$} 	&  \colhead{mK} 		&  \colhead{km s$^{- 1}$} 
 }
 \startdata
 J 1 $\rightarrow$ 0 		& 22 48106 			& 1.079			& 8.999 					& -27(1)\tablenotemark{c}	& 13.4(1)\tablenotemark{c}& $\geq$ -7				& \nodata				& \nodata					& \nodata				& $\geq$ -17			& \nodata				\\
 J 2 $\rightarrow$ 1 		& 44 979.54 			& 3.238			& 18.001 					& -70(2)\tablenotemark{c}	& 14.7(7)\tablenotemark{c}& $\geq$ -9				& \nodata				& \nodata					& \nodata				&  -46(18)\tablenotemark{c}				& 3(1)\tablenotemark{c}				\\
 \\
 J 6 $\rightarrow$ 5 		& 134 932.73 			& 22.665			& 53.998 					& $\leq$52			& \nodata				& $\leq$ 71				& \nodata				& 28\tablenotemark{b}		& 9\tablenotemark{a}	& \nodata 				& \nodata				\\ 
 J 7 $\rightarrow$ 6 		& 157 418.72			& 30.220			& 63.002 					& 99\tablenotemark{b}	& \nodata				& $\leq$ 34				& \nodata				& 34\tablenotemark{b}		& 9\tablenotemark{a}	& \nodata				& \nodata				\\ 
 \enddata
 \label{observedtransitions}
 \tablenotetext{a}{Velocity width taken from Pulliam et al. (2012)}
 \tablenotetext{b}{Blended}
 \tablenotetext{c}{Results of Gaussian fits to the observations with 1$\sigma$ uncertainties given in units of the last significant figure.}
 \end{deluxetable}
 
 \begin{deluxetable}{c c c c c c c c c c c c}
\tablewidth{0pt}
\tablecaption{Transitions of $l$-C$_3$H observed toward Sgr B2(N) and TMC-1.}
\tablecolumns{12}
\tabletypesize{\footnotesize}
\tablehead{
\multicolumn{3}{c}{} 																		& \colhead{} 							& \colhead{} 		& \colhead{} 				& 							\multicolumn{4}{c}{Sgr B2(N)} 													& \multicolumn{2}{c}{TMC-1\tablenotemark{a}} 	\\ \cline{7-12}
\multicolumn{3}{c}{} 																		& \colhead{} 							& \colhead{} 		& \colhead{} 				& \multicolumn{2}{c}{+64 km s$^{-1}$} 					& \multicolumn{2}{c}{+82 km s$^{-1}$} 	&  \colhead{} 			&  \colhead{} 								\\ \cline {7-10}
\multicolumn{3}{c}{Transition} 																& \colhead{Frequency\tablenotemark{a}}	 	& \colhead{E$_u$} 	& \colhead{} 				& \colhead{T$_A^*$}	&  \colhead{$\Delta V$} 			&  \colhead{T$_A^*$} 				&  \colhead{$\Delta V$} 	&  \colhead{T$_A^*$} 	&  \colhead{$\Delta V$} 	\\ \cline{1-3}
\colhead{$J^{\prime}-J^{\prime\prime}$}	& \colhead{Parity}	& \colhead{$F^{\prime}-F^{\prime\prime}$}	& \colhead{MHz} 						& \colhead{K} 		& \colhead{S$_{ij}\mu^2$} 	& \colhead{mK} 		&  \colhead{km s$^{-1}$} 			&  \colhead{mK}					&  \colhead{km s$^{-1}$} 	&  \colhead{mK} 		&  \colhead{km s$^{- 1}$} 
 }
 \startdata
$3/2 - 1/2$						& $f$				& $1 - 1$								& 32 617.016							& 1.56622			& 4.189					& -22\tablenotemark{c}		& \nodata\tablenotemark{c}	& $\geq$ -7						& \nodata				& 78					&	0.39				\\
								& 				& $2 - 1$								& 32 627.297							& 1.56672			& 20.932					& -88(1)					& 11.5(2)					& -19							& \nodata				& 287				&	0.47				\\
								& 				& $1 - 0$								& 32 634.389							& 1.56619			& 8.370					& -59(2)					& 14.6(7)					& -42							& \nodata				& 96					&	0.75				\\
								\\
								& $e$			& $2 - 1$								& 32 660.645							& 1.56990			& 20.932					& -60(2)					& 9.0(3)						& -15							& \nodata				& 251				&	0.47				\\
								& 				& $1 - 0$								& 32 663.361							& 1.57032			& 8.372					& -9\tablenotemark{b,c}		& \nodata	\tablenotemark{c}	& $\geq$ -7						& \nodata				& 99					&	0.43				\\
								& 				& $1 - 1$								& 32 667.668							& 1.57024			& 4.184					& -22\tablenotemark{c}		& \nodata	\tablenotemark{c}				& $\geq$ -7						& \nodata				& 61					&	0.48				\\
 \enddata
 \label{neutraltransitions}
 \tablecomments{Except where noted, values of T$_A^*$ and $\Delta V$ for the Sgr B2(N) +64 km s$^{-1}$ data were obtained by Gaussian fits with 1$\sigma$ uncertainties given in units of the last significant digit.  In the case of Sgr B2(OH), no fits were performed, and T$_A^*$ is listed either as peak intensity or as an RMS noise level.}
 \tablenotetext{a}{Values from Kaifu et al. (2004)}
 \tablenotetext{b}{Affected by local, non-zero baseline}
 \tablenotetext{c}{Unable to fit a Gaussian - T$_A^*$ taken as peak intensity, no linedwidth determined}
 \end{deluxetable}

\section{Results}
\label{data}

Following the convention of Remijan et al. (2005), the total column density ($N_T$) of a species observed in emission is given by
\begin{equation}
\label{emissioncd}
N_T=1.8\times10^{14}\times\frac{Q_re^{E_u/T_{ex}}}{\nu S\mu ^2}\times\frac{\Delta T_A^* \Delta V/\eta _B}{1-\frac{(e^{(4.8\times10^{-5})\nu /T_{ex}}-1)}{(e^{(4.8\times10^{-5})\nu /T_{bg}}-1)}}\mbox{ cm}^{-2}
\end{equation}
while for absorption, the relationship becomes
\begin{equation}
\label{absorptioncd}
N_T=8.5\times10^9\times\frac{Qr(\Delta T_A^*\Delta V/\eta _B)}{(T_{ex}-T_c/\eta _B)S\mu ^2(e^{-E_l/T_{ex}}-e^{-Eu/T_{ex}})}\mbox{ cm}^{-2}
\end{equation}
where line shapes are assumed to be Gaussian, $\eta_B$ is the telescope beam efficiency, $T_{ex}$ is the excitation temperature (K), $T_{bg}$ is the background temperature (K), $T_c$ is the continuum temperature (K), $Q_r$ is the rotational partition function, $E_u$ and $E_l$ are the upper and lower state energies, respectively (K), $\Delta T_A^*$ is the line intensity (mK), $\Delta V$ is the line width (km s$^{-1}$), $S\mu^2$ is the product of the line strength and square of the dipole moment (debye$^2$), and $\nu$ is the transition frequency (MHz).

\subsection{Sgr B2(N)}

Using Equation \ref{absorptioncd}, a best fit value of $T_{ex}\simeq10$ K is found for the transitions observed towards Sgr B2(N), giving a total column density of B11244 of $\sim$$10^{13}$ cm$^{-2}$. Linewidths of 13.4 and 14.7 km$^{-1}$ were assumed based on Gaussian fits to the absorption profiles, and a dipole moment of $\mu=$ 3 Debye was used following Pety et al. (2012).  Predicted line profiles for the these transitions are shown as dashed blue lines in Figure \ref{b2n} using the derived values for $T_{ex}$ and column density.   The observations of B11244 absorption in Sgr B2(N) indicate the signal likely arises from cold, diffuse gas surrounding the hot, dense core, rather than from the hot core itself, consistent with observations of other cold, extended species in this source (see e.g. Neill et al. (2012) \& Hollis et al. (2004)).  The derived abundance is similar to those derived in previous observations of small organic molecules toward this source \citep{Nummelin2000}.

For comparison, we calculated an approximate column density of neutral $l$-C$_3$H using the transitions shown in Figure \ref{linear} from PRIMOS and Equation \ref{absorptioncd}.  Although six transitions of neutral $l$-C$_3$H are observed, all are hyperfine components of the single $J = 3/2 - 1/2$ manifold.  As a result, two parameter ($T_{ex}$ and $N_T$) fits are not well-constrained.  We have therefore proceeded on the assumption that, as they are both observed in absorption and are likely co-spatial, neutral $l$-C$_3$H and B11244 will be similar in excitation temperature.  We find a value of $T_{ex}\simeq8.7$ K with a column density of $\sim$$10^{14}$ cm$^{-2}$ provides a good approximation.  This results in a ratio of neutral:B11244 in Sgr B2(N) of $\sim$6:1, consistent with the ratio  Pety et al. (2012) derive of $\sim$4:1 in the Horsehead PDR.   High-resolution maps of both neutral $l$-C$_3$H and B11244 would greatly aid in determining the validity of these assumptions.

The observed behavior of B11244 in moving from absorption to emission with increasing frequency is not unique to this molecule in Sgr B2(N).  For example, CH$_2$OHCHO \citep{Hollis2004}, CNCHO \citep{Remijan2008A}, CH$_3$CHO \citep{Hollis2006} and especially HCCCHO (R. Loomis, private communication) among others (including CH$_3$OH and H$_2$CO), display similar behavior.  This is largely a function of decreasing continuum temperature with frequency.  In fact, the $J=6-5$ and $J=7-6$ transitions provide a constraint on the background continuum temperatures above the CMB of $\sim$1 K at 135 GHz and $\sim$0.5 K at 157 GHz.

\subsection{Sgr B2(OH)} 

In Sgr B2(OH), only two lines fall within the frequency of the Turner Survey observations, the lower of which (at 135 GHz) is clearly blended (see Figure \ref{b2oh}).  As such, here we calculate only an upper limit to B11244 column density on the assumption that all of the emission at the peak of each signal arises from B11244.  A line width of 9 km s$^{-1}$ was assumed based on the analysis of nearby spectral regions from the same dataset by Pulliam et al. (2012).  Based on these values, a best fit $T_{ex}$ of $\sim$79 K gives an upper limit on the column density of $\leq1.5\times10^{13}$ cm$^{-2}$.  

\subsection{TMC-1}

The lack of any definitive signal from B11244 towards TMC-1 precludes any quantitative determination of a column density.  However, assuming the weak absorption signal at 44.9 GHz does arise from  $J = 2 - 1$ of B11244, a zeroeth-order approximation of the column density can be obtained using Equation \ref{absorptioncd} and an estimated temperature of $\sim$9 K \citep{Kalenskii2004}.  Such an analysis results in an estimated upper limit to the column density of $\sim$$6 \times 10^{11}$ cm$^{-2}$.

For comparison, the column density of neutral $l$-C$_3$H, assuming the same temperature of $\sim$9 K, is $\sim$$9 \times 10^{12}$ cm$^{-2}$ in this source.  This results in a ratio of 15:1, two and a half times that in Sgr B2(N) and almost four times that in the Horsehead PDR.  Given the large uncertainties involved, however, these numbers are not inconsistent.

\section{Spectral Fitting}
\label{fitting}

The spectroscopic parameters and line list for $l$-C$_3$H$^+$ as listed in the CDMS catalog is accessible in full via www.splatalogue.net.  Huang et al. (2013) question the assignment of the observed transitions of Pety et al. (2012) to $l$-C$_3$H$^+$ based on large discrepancies between observed and calculated values for the $D$ and $H$ distortion constants.  The predictions of Pety et al. (2012) were robust enough to predict the $J = 1-0$ and $J = 2-1$ transitions presented here. However, in an effort to confirm their values for the $D$ and $H$ constants, we have refit the molecular signals using the frequencies for the $J = 1-0$ and $J = 2-1$ transitions determined from our observations.

The observed transitions of B11244 were fit using the CALPGM \citep{Pickett1991} program suite, using a standard linear rotor Hamiltonian giving energies as shown in Equation \ref{Hamiltonian}.
\begin{equation}
\label{Hamiltonian}
E(J) =  BJ(J+1) - D(J(J+1))^2 + H(J(J+1))^3 + L(J(J+1))^4 + M(J(J+1))^5 + (...)
\end{equation}
CALPGM is designed to fit spectroscopic constants of a model Hamiltonian to a set of observed transitions using an iterative least-squares fitting algorithm. Following Pety et al. (2012), we include first ($D$) and second ($H$) order centrifugal distortion constants in the Hamiltonian producing a fit to the unshifted transitions with an RMS observed minus calculated value of 0.0862 MHz, below the average observational uncertainty of 0.1015 MHz.  Addition of higher order terms, $L$ and $M$ in equation \ref{Hamiltonian} are found to improve the RMS of the fit by $\sim$10 kHz for each additional term and give a RMS change in the predicted frequencies of $\sim$90 kHz and 80 kHz for $L$ and $M$ respectively. These corrections are below the experimental uncertainty and therefore of questionable physical significance, thus only the first and second order corrections are included in the final fits. As a final step the 1$\sigma$ uncertainty of all fit parameters was computed using the PIFORM program.\footnote{Z. Kisiel, PIFORM available at http://www.ifpan.edu.pl/$\sim$kisiel/asym/asym.htm\#piform} 

The spectral resolution of the PRIMOS observations is significantly higher than those from Pety et al. (2012); $\sim$21 kHz versus 49 and 195 kHz.  However, the linewidths of the observed transitions in Sgr B2(N) used in the fit are broad compared to those observed by Pety et al. (2012) in the Horsehead PDR (13.3 \& 14.7 km s$^{-1}$ versus $\sim$1 km s$^{-1}$), thus introducing additional uncertainty in the measurement of the line centers.  Further, as discussed by Pety et al. (2012), the absolute accuracy of the spectroscopic constants is determined by the accuracy to which the $V_{LSR}$ velocity of B11244 emission is known.  The use of two independent observational datasets towards two separate sources compounds this issue.

While we cannot mitigate the uncertainty due to the broad lineshapes, we have attempted to minimize the effects of uncertainty in $V_{LSR}$.  To account for this, a minimization of the fit RMS was performed by varying the $V_{LSR}$ offsets for the PRIMOS dataset and the IRAM dataset independently over a range of -3 km s$^{-1}$ to +3 km s$^{-1}$.  The results of this analysis indicate convergence is quickly reached with minimal variation in either dataset.  Based on their observations of similar molecules, Pety et al. (2012) determine an uncertainty level of $\pm$0.2 km s$^{-1}$ for the IRAM dataset.  Under these constrains, a minimum RMS is achieved with offsets to the PRIMOS dataset of 0.4 - 0.8 km s$^{-1}$ - equivalent to approximately twice the resolution of the observations at 22 GHz.  For comparison, a PRIMOS offset of 0.0 km s$^{-1}$ requires an IRAM offset of -0.6 km s$^{-1}$ for minimization.  

The best fit rotational constants found at the outer limits of the best fit region (assuming $\pm$0.2 km s$^{-1}$ offsets to the IRAM data) are shown in Table \ref{calpgm}.  The absolute variance in the $B$ rotational constant at the outer limits of the IRAM offsets is found to be 15 kHz, much less than the resolution of the observations.   The implications of these results are discussed in the following section.

\begin{deluxetable}{l c r c l c r }
\tablewidth{0pt}
\tablecaption{Best-fit spectroscopic constants obtained by shifting V$_{LSR}$ for the PRIMOS and IRAM datasets.}
\tablecolumns{7}
\tablehead{
\multicolumn{3}{c}{$\Delta$V$_{LSR}$ (km s$^{-1}$)}		&&			 \multicolumn{3}{c}{$\Delta$V$_{LSR}$ (km s$^{-1}$)} \\
\multicolumn{3}{c}{IRAM -0.2, PRIMOS 0.4}		&&			 \multicolumn{3}{c}{IRAM 0.2, PRIMOS 0.8} }
\startdata
B	   &    11244.9421(41)   &  MHz			&&	B	   &    11244.9571(41)   &  MHz			\\
D	   &    7.745(80)	     &  kHz				&&	D	   &    7.745(80)	     &  kHz				 \\
H	   &    0.49(37)	     &  Hz				&&	H	   &    0.49(37)	     &  Hz				\\
Fit RMS	   &    31.9 	     &  kHz				&&	Fit RMS	   &    31.9 	     &  kHz				 \\
\hline
\enddata
\label{calpgm}
\tablecomments{1$\sigma$ uncertainties on spectroscopic constants (type A, k = 1 \citep{Taylor1994}) are given in parentheses in units of the last significant digit}
\end{deluxetable}

\section{Discussion}

At first glance, the presence of $l$-C$_3$H$^+$ would be remarkable, as the species is known to react readily (and destructively) with H$_2$ \citep{Anicich1986}.  The arguments for the assignment of these features to $l$-C$_3$H$^+$ by Pety et al. (2012), however, appear robust.  $l$-C$_3$H$^+$ is thought to be a key intermediate in the production of small hydrocarbon molecules, including $l$-C$_3$H. Indeed, the detected abundance of $l$-C$_3$H$^+$ in the Horsehead PDR is remarkably consistent with chemical models of the region performed by Pety et al. (2012). Additionally, as discussed by Pety et al. (2012), the reaction rate of the destructive reaction of $l$-C$_3$H$^+$ with H$_2$ is strongly dependent on the gas temperature, with very low temperatures ($T < 20$ K) and especially warmer temperatures ($T > 50$ K) decreasing the reaction rate coefficients \citep{Savic2005}.  Therefore, perhaps it is not surprising to have found cold $l$-C$_3$H$^+$ ($T < 11$ K) in Sgr B2(N) and possibly TMC-1, but warm ($T \sim$80 K) $l$-C$_3$H$^+$ in Sgr B2(OH).

Due to the challenges discussed in \S\ref{fitting}, the constants we determine from our spectroscopic fit have greater uncertainties than those determined by Pety et al. (2012).  These may in fact be more a faithful reflection of the true uncertainties than those presented by Pety et al. (2012), as our fit takes into account the inherent uncertainties in $V_{LSR}$.  Nevertheless, our analysis agrees quite well with that of Pety et al. (2012); we find the values for $D$ and $H$ do not vary from theirs within the stated uncertainties.  Thus, we conclude that the fit presented by Pety et al. (2012) is a faithful representation of the detected molecular signatures and is consistent with a closed-shell, linear molecule.  Further, the abundances and physical conditions are consistent with the current understanding of $l$-C$_3$H$^+$ and $l$-C$_3$H chemistry.

Resolving the discrepancies presented by Huang et al. (2013) through astronomical observations will certainly require further, higher-frequency observations of the molecule.  As the effects of the $D$ and $H$ constants become exponentially more pronounced with higher J-levels, each additional line measured beyond those found by Pety et al. (2012) will serve to lock these values into place.  Indeed, by 315 GHz, the difference in the predicted line frequencies using the $D$ and $H$ constants of Pety et al. (2012) and Huang et al. (2013) differ by more than 9 km s$^{-1}$.  Thus, observation of these lines in Sgr B2(OH), where the warmer conditions favor lines in this frequency range, could help to resolve this issue despite the broad linewidths observed there.

There is, however, no substitute for laboratory data, and although further astronomical observations could certainly help to resolve the question, they cannot approach the level of confidence found in experiments in a laboratory setting.  Thus, laboratory measurements using absolute frequency standards and controlled production conditions are warranted to expand the spectroscopic study of $l$-C$_3$H$^+$.  The laboratory observation of small hydrocarbon and hydrocarbon chain neutrals, cations, and anions is a well-established, if non-trivial process (see e.g. McCarthy et al. (2006), Bogey et al. (1984)).  

\section{Conclusions}

Here, we have presented observations of the $J=1-0$ and $J=2-1$ transitions of the B11244 molecule in Sgr B2(N), and observations of the $J=6-5$ and $J=7-6$ transitions in Sgr B2(N) and Sgr B2(OH) using the publicly available PRIMOS data and the Barry E. Turner Legacy Survey.  Neutral $l$-C$_3$H has been detected in Sgr B2(N) in a ratio consistent with that found in the Horsehead PDR.  Observations of TMC-1 reveal strong $l$-C$_3$H signals and a tentative detection of a weak B11244 transition.  A spectroscopic fit of the molecule including the newly observed $J=1-0$ and $J=2-1$ transitions agrees with that of Pety et al. (2012), but does not resolve the discrepancy with the calculated constants of Huang et al. (2013).  Follow-up observational and laboratory studies are warranted to definitively identify the molecule.

\acknowledgments

The authors are grateful to M. Ohishi for providing the observational data towards TMC-1 and to the anonymous referee for very helpful comments.  B.A.M. gratefully acknowledges funding by an NSF Graduate Research Fellowship.  The National Radio Astronomy Observatory is a facility of the National Science Foundation operated under cooperative agreement by Associated Universities, Inc.

\end{document}